\documentclass{article}

\usepackage{microtype}
\usepackage{graphicx}
\usepackage{subfigure}
\usepackage{booktabs} 
\usepackage{multirow}
\usepackage{multicol}
\usepackage{bbm}
\usepackage{hyperref}

\usepackage[accepted]{icml2025}

\usepackage{amsmath}
\usepackage{amssymb}
\usepackage{mathtools}
\usepackage{amsthm}
\usepackage{bm}

\usepackage[capitalize,noabbrev]{cleveref}

\usepackage{enumitem}

\theoremstyle{plain}

\theoremstyle{definition}

\theoremstyle{remark}

\usepackage[textsize=tiny]{todonotes}

\icmltitlerunning{Simulation-Based Pretraining and Domain Adaptation for Astronomical Time Series}

\begin{document}

\twocolumn[
\icmltitle{Simulation-Based Pretraining and Domain Adaptation for Astronomical Time Series with Minimal Labeled Data}




\begin{icmlauthorlist}
\icmlauthor{Rithwik Gupta}{mit,irvington}
\icmlauthor{Daniel Muthukrishna}{mit,cfa,astroai}
\icmlauthor{Jeroen Audenaert}{mit}

\end{icmlauthorlist}

\icmlaffiliation{mit}{Massachusetts Institute of Technology, Cambridge, MA 02139, USA}
\icmlaffiliation{irvington}{Irvington High School, Fremont, CA 94538, USA}
\icmlaffiliation{cfa}{Center for Astrophysics, Harvard \& Smithsonian, Cambridge, MA 02138, USA}
\icmlaffiliation{astroai}{AstroAI}


\icmlcorrespondingauthor{Daniel Muthukrishna}{danmuth@mit.edu}
\icmlcorrespondingauthor{Rithwik Gupta}{rithwikca2020@gmail.com}

\icmlkeywords{Machine Learning}

\vskip 0.3in
]

\printAffiliationsAndNotice

\begin{abstract}
Astronomical time-series analysis faces a critical limitation: the scarcity of labeled observational data. We present a pre-training approach that leverages simulations, significantly reducing the need for labeled examples from real observations. Our models, trained on simulated data from multiple astronomical surveys (ZTF and LSST), learn generalizable representations that transfer effectively to downstream tasks. Using classifier-based architectures enhanced with contrastive and adversarial objectives, we create domain-agnostic models that demonstrate substantial performance improvements over baseline methods in classification, redshift estimation, and anomaly detection when fine-tuned with minimal real data. Remarkably, our models exhibit effective zero-shot transfer capabilities, achieving comparable performance on future telescope (LSST) simulations when trained solely on existing telescope (ZTF) data. Furthermore, they generalize to very different astronomical phenomena (namely variable stars from NASA's \textit{Kepler} telescope) despite being trained on transient events, demonstrating cross-domain capabilities. Our approach provides a practical solution for building general models when labeled data is scarce, but domain knowledge can be encoded in simulations.

\end{abstract}

\vspace{-2em}

\section{Introduction}
Time-series analysis in astronomy often requires substantial labeled data for supervised learning approaches. Models have been developed to classify variable stars and transient events \citep[e.g.][]{Muthukrishna19RAPID, Hlozek2023-Plasticc-results, Audenaert2025}, detect anomalies \citep{vraenn, Muthukrishna2022, Perez-Carrasco_2023}, and estimate physical parameters such as redshift \citep[e.g][]{redshift-transients, maven}. However, these models are typically trained for specific telescopes or individual tasks, with limited ability to transfer learned representations across different instruments. Furthermore, they require extensive labeled data from new telescopes to achieve adequate performance, limiting their effectiveness during the initial survey stages. 

Foundation models (FMs) have transformed natural language processing and computer vision by learning generalizable representations from large quantities of data \citep{bommasani2022opportunitiesrisksfoundationmodels}. However, astronomical data presents unique challenges: (1) limited publicly available labeled data; (2) instrument-specific characteristics that hinder cross-survey generalization \citep{CausalFM}; and (3) complex physical phenomena that require domain expertise to model effectively. Recent work has begun to develop foundation models for astrophysical data \citep[e.g.][]{AstroCLIP,astromer, smith2024astroptscalinglargeobservation,maven,CausalFM}. While these approaches demonstrate cross-modal learning, no existing approach achieves zero-shot transfer between different telescopes for time-domain phenomena. Moreover, existing FMs typically use self-supervised methods that do not take advantage of the astronomical class structure that is fundamental to astronomical understanding.

\begin{figure*}
    \centering
    \includegraphics[width=\linewidth]{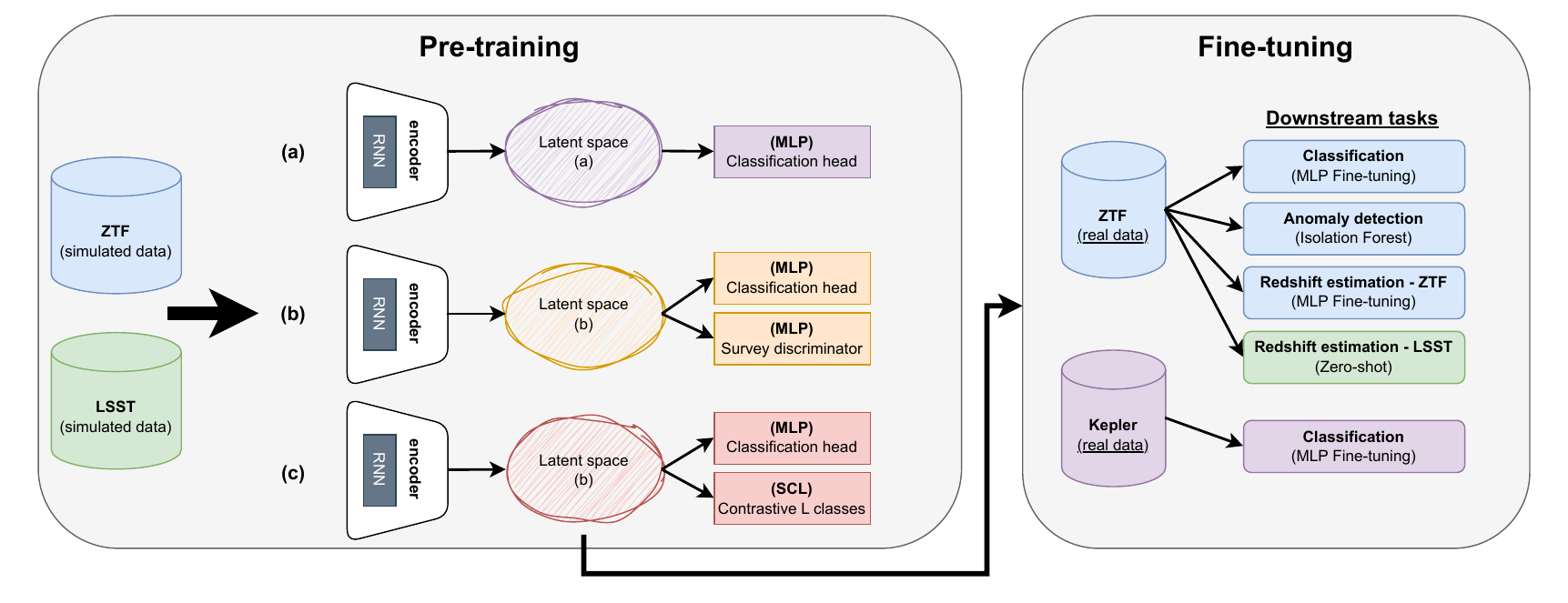}
    \vspace{-2em}
    \caption{Overview of our simulation-based pre-training methodology showing three model variants (Classifier, Adversarial, Contrastive) and evaluation on both in-domain and cross-domain tasks. We first pre-train various classifiers and domain-agnostic models using simulated data. We then evaluate these models on various downstream tasks, including zero-shot estimation for new telescopes.}
    \label{fig:architecture}
\end{figure*}

Astronomy has a significant advantage over many domains: decades of physical understanding encoded in simulations. Astronomers have developed detailed models of astrophysical phenomena that generate synthetic light curves \citep[e.g.][]{plasticc_modelers_2019_2612896}. While these simulations contain systematic differences from real observations, they capture the underlying physical processes and can provide effective pretraining for transfer to real data \citep{gupta2025}.

Motivated by the success of supervised classifier-based methods for anomaly detection \citep{gupta2024}, we propose a novel approach to learning generalizable representations for astronomical time series that:
\begin{enumerate}[noitemsep,topsep=0pt]
    \item Leverages the latent space of classifiers pretrained on simulations from multiple astronomical surveys
    \item Develops domain-agnostic representations through adversarial and contrastive learning objectives
    \item Enables effective downstream task performance with minimal labeled real data
    \item Demonstrates zero-shot transfer capabilities between different survey telescopes
\end{enumerate}

Unlike existing approaches that focus on single-survey applications, our method specifically targets cross-survey transfer learning using physics-informed simulations.

This approach is particularly valuable for upcoming surveys like the Vera C. Rubin Observatory's Legacy Survey of Space and Time (LSST), which will produce millions of time-series alerts nightly \citep{Ivezic2009LSST:Products}. Having models ready to analyze LSST data from day one---without requiring extensive new labeled datasets---would dramatically accelerate scientific discovery.\footnote{The code used in this work is publicly available: \hyperlink{https://github.com/Rithwik-G/astrofm2.0}{https://github.com/Rithwik-G/astrofm2.0} and \hyperlink{https://github.com/Rithwik-G/Kepler-FM}{https://github.com/Rithwik-G/Kepler-FM}}

\section{Datasets and Benchmarks}
\label{sec:datasets}

\subsection{Pretraining Data}
We pretrain our models using 151,468 simulated astronomical transients from two survey telescopes: 87,080 from the Zwicky Transient Facility (ZTF; \citealp{ZTF}) and 64,388 from LSST. We evaluate three distinct pretraining configurations: (1) ZTF simulations only, (2) combined ZTF and LSST simulations, and (3) combined ZTF and LSST simulations with domain-agnostic objectives (adversarial or contrastive training).

The simulated transients encompass eight astronomical classes: Type Ia, Ib/c, II, Iax, and Ia-91bg supernovae; Tidal Disruption Events; Super-Luminous Supernovae Type I; and Active Galactic Nuclei (see Table~\ref{table:simulated} for class distributions). 

ZTF simulations use established astrophysical models \citep{PlasticcSim, Muthukrishna19RAPID} reflecting the currently operational survey's characteristics including two photometric bands ($gr$) with limiting magnitude $r \sim 21$ mag. LSST simulations \citep{elasticc} model the Rubin Observatory beginning operations in late 2025: six photometric bands ($ugrizy$) with significantly deeper sensitivity ($r \sim 25$ mag).

Each transient is represented as a multichannel variable-length time series (light curve\footnote{See Figure \ref{fig:architecture} for a plot of example light curves}) with observations formatted as $[\lambda_p, t_i, f_i]$, where $\lambda_p$ indicates the median passband wavelength, $t_i$ is the time since the first observation (in days), and $f_i$ is the normalized flux (brightness measurement). This standardized representation enables cross-survey compatibility by encoding temporal and spectral information in a telescope-agnostic format. 

\subsection{Evaluation Datasets and Downstream Tasks}
We evaluate pretrained models by fine-tuning on downstream tasks using real observational data and additional simulated datasets to assess in-domain performance, cross-domain generalization, and zero-shot transfer capabilities.

\paragraph{ZTF Real Data:} 3,747 spectroscopically confirmed transients from the ZTF Bright Transient Survey \citep{Rehemtulla_2024} spanning three task types:
\begin{itemize}[topsep=-3pt]
    \item \textit{Classification}: Macro-averaged Area Under the ROC Curve (AUROC) for distinguishing Type Ia (771 objects), Type Ib/c (2,828 objects), and Type II (148 objects) supernovae
   \item \textit{Redshift Estimation}: Mean Squared Error (MSE) for predicting spectroscopic redshifts using photometric light curve data from the same 3,747 objects
    \item \textit{Anomaly Detection (AD)}: Binary classification AUROC treating 38 rare transient types as anomalies. The models are not provided any anomalous data during training.
\end{itemize}

\paragraph{Simulated Redshift Estimation:} 10,000 ZTF and 9,990 LSST simulated objects with redshift labels for regression tasks (MSE evaluation). These datasets are distinct from the pretraining classification data and the pretrained models have no prior exposure to redshift information and must learn this novel regression task during fine-tuning. For zero-shot evaluation, we use a subset of 2,596 LSST objects restricted to ZTF's redshift range ($\mathrm{z} < 0.4$), where models fine-tuned exclusively on ZTF redshift data predict LSST redshifts without any LSST-specific training redshifts.

\paragraph{Kepler Real Data (Cross-Domain):} 9,501 variable stars observed by NASA's Kepler telescope \citep{Koch2010,Borucki2010} and compiled by \citet{Audenaert2021}, spanning nine stellar variability classes (aperiodic, constant, contact binaries, $\delta$ Scuti / $\beta$ Cephei, eclipsing binaries, $\gamma$ Doradus / Slowly Pulsating B stars, instrument variability, RR Lyrae / Cepheids, and Solar-like oscillators) to test generalization from transient to variable star \textit{classification} (macro-averaged AUROC). Kepler is a space telescope designed for high-cadence (30-minute) photometric monitoring in a single broad optical band, contrasting sharply with the multi-band, ~3-day cadence observations of ground-based transient surveys like ZTF and LSST.

\subsection{Evaluation Scenarios}
To simulate realistic deployment challenges when applying pretrained models to new surveys with varying data availability, we evaluate under three scenarios:

\begin{description}[topsep=-3pt,noitemsep]
    \item[Limited Scenario:] Fine-tunes pretrained models using only 512 randomly selected labeled objects per task. We repeat experiments with five different random samples to account for selection variance.
    \item[Full Scenario:] Fine-tunes pretrained models using all available labeled data for each task, representing performance after extensive labeling efforts.
    \item[Zero-Shot Scenario:] Applied exclusively to LSST redshift estimation, where models fine-tuned only on ZTF redshift data are directly applied to LSST simulations without any LSST-specific fine-tuning data. This tests immediate model deployment capability on new surveys before domain-specific labeling becomes available.
\end{description}
The zero-shot redshift evaluation deliberately restricts LSST objects to ZTF's redshift range because our goal is leveraging existing telescope knowledge for immediate application to new surveys. While LSST will observe significantly higher redshifts than ZTF (see Figure 2), zero-shot models cannot reasonably extrapolate beyond their training distribution.

\section{Methods}

We develop three simulation-based pretraining approaches for learning generalizable representations of astronomical time series: a classifier trained on ZTF and LSST simulations, and a classifier enhanced with domain-agnostic objectives using either an adversarial objective or a contrastive objective for explicit survey alignment.

\subsection{Base Classifier}

Neural network classifiers have demonstrated the ability to learn meaningful representations of astronomical phenomena that transfer effectively to tasks beyond their original classification objective, including anomaly detection and morphological analysis \citep{WalmsleyCNN, etsebeth2023astronomaly, Gupta2025-anomalydetection}. Building on these successes, our base model trains a classifier on simulated transients from both ZTF and LSST to distinguish between the eight astronomical transient classes. 

\textbf{Architecture:} We employ a Gated Recurrent Unit (GRU; \citealt{GRU}) network that processes variable-length time series inputs. Each observation $[\lambda_p, t_i, f_i]$ is processed sequentially through the 100-unit GRU layers, with the final hidden state passed through two fully-connected mutlilayer perceptron (MLP) layers to produce class predictions. The model is trained using a standard categorical cross-entropy loss $H$ on the final softmax predictions. We extract the penultimate MLP layer as our learned latent representation $ \bm{z}_i \in \mathbb{R}^{128}$
for downstream tasks. We refer to the classifier with its final classification layer removed as the encoder $C_L$ throughout this work, such that $\bm{z}_i = C_L(\bm{X}_i)$ for each input light curve $\bm{X}_i$.

While this classifier learns structured representations that distinguish astronomical classes within each survey (see Fig.\,4 of \citealp{gupta2024}), it does not produce a telescope-agnostic latent space. Objects from the same class but different surveys remain separated in the latent space (see Figures \ref{fig:umap} and \ref{fig:umapclass} for UMAP visualizations of the latent space).

\subsection{Adversarial Training}

Different telescopes have distinct observational characteristics (photometric bands, sensitivity, cadence, noise) that can cause models to learn survey-specific artifacts rather than the underlying physics. To create domain-agnostic representations that work across different astronomical surveys, we implement an adversarial training framework that encourages the model to learn telescope-invariant features useful for astronomical classification.

\textbf{Architecture}: We extend the base classifier with an additional discriminator network $D$ that attempts to predict whether a representation came from ZTF or LSST simulations. The discriminator is a two-layer MLP that takes the latent representation $\bm{z}_i$ as input to predict the survey telescope.

\textbf{Training}: The classifier $C(\bm{X}_i)$ and the discriminator $D(\bm{z}_i)$ are trained simultaneously with competing objectives, 
\begin{align}
    \textbf{Classifier loss:} \quad &\mathcal{L}_{C} = H(C(\bm{X}_i), c_i) \\
    \textbf{Discriminator loss:} \quad &\mathcal{L}_{D} = H(D(\bm{z}_i), O_i)
\end{align}
where $H$ denotes the cross-entropy loss, $c_i$ is the class label, and $O_i \in \{\text{ZTF}, \text{LSST}\}$ indicates the source survey.
The model is trained to both classify the transient correctly and to confuse the discriminator, thereby learning a latent space that is useful for classification yet agnostic to the survey domain. The adversarial training process is detailed further in Appendix \ref{sec:adv_training}.

\subsection{Supervised Contrastive Training}

While adversarial training encourages domain-agnostic representations, it does not explicitly enforce that objects of the same class cluster together across surveys. Hence, as a comparative method, we use supervised contrastive learning \citep{SCL}, which explicitly pulls together representations of the same class from different surveys while pushing apart different classes. We use a variant of the contrastive loss proposed in \citet{SimCLR} for model pretraining where we treat all samples belonging to the same class as positive pairs. This modification provides a more explicit signal for class-based alignment and naturally encourages unification across domains, as long as examples from the same class are drawn from both surveys. The total loss is the sum of the classifier loss $\mathcal{L}_C$ and the supervised contrastive loss $\mathcal{L}_{\mathrm{SCL}}$. The training details and algorithm is described in Appendix \ref{sec:con_training}.

\subsection{Baselines}
We compare our pretrained models against models trained directly on downstream tasks without pretraining, which reflects the standard approach in prior work.
We also train classifiers on individual surveys (notably ZTF-only) as additional baselines to isolate the benefits of multi-survey pretraining.

\subsection{Downstream Tasks}
For downstream tasks, we freeze the pretrained encoder and attach task-specific multi-layer perceptrons (MLPs). This preserves the learned cross-survey representations while adapting to specific objectives (classification, regression, anomaly detection). The fine tuning for each task is set up as follows (see Appendix \ref{sec:fine-tuning} for further details):
\begin{description}
    \item[Classification/Redshift estimation:] Two-layer MLP with appropriate output activation (softmax for classification, linear for regression).
    \item[Anomaly Detection:] Following \citet{Gupta2025-anomalydetection}, we train an isolation forest on frozen latent representations, treating rare transient classes as anomalies
    \item[Cross-Domain Tasks (Kepler):] For novel domains, we unfreeze the encoder layers to allow limited adaptation while preserving the core learned representations
    \item[Zero-Shot Transfer:] For zero-shot evaluation on LSST redshift estimation, we use a two-layer MLP to fine-tune the pretrained models on ZTF redshift data only, and then evaluate on LSST simulations without any LSST training data. We test two approaches: direct MLP prediction and kNN (k=100) using distance-weighted averaging of the closest ZTF embeddings.
\end{description}

\section{Results}

We evaluate our simulation-based pretraining models on multiple downstream tasks using real observational data from ZTF (3,747 transients) and Kepler (9,501 variable stars), as well as simulated redshift estimation data from ZTF (10,000 objects) and LSST (9,990 objects). We compare both using the ``Full'' dataset for fine-tuning and using only a ``Limited'' set of 512 objects. Our experiments demonstrate that pretraining on simulations provides substantial improvements over training from scratch (no pretraining), with particularly pronounced benefits for cross-survey generalization and zero-shot transfer.

\begin{table*}
    \centering
    \resizebox{0.7\textwidth}{!}{
    \small
    \begin{tabular}{l@{\hspace{6pt}}|cc@{\hspace{8pt}}|cc@{\hspace{8pt}}|cc@{\hspace{8pt}}|c@{\hspace{8pt}}|cc}
        \toprule
        \multirow{3}{*}{\textbf{Model}} & \multicolumn{6}{c|}{\textbf{ZTF Real Data}} & \textbf{Kepler} & \multicolumn{2}{c}{\textbf{Simulations}} \\
        \cmidrule(lr){2-7} \cmidrule(lr){8-8} \cmidrule(lr){9-10}
        & \multicolumn{2}{c|}{\textbf{Classification}} & \multicolumn{2}{c|}{\textbf{Redshift}} & \multicolumn{2}{c|}{\textbf{Anomaly Det.}} & \textbf{Classification} & \multicolumn{2}{c}{\textbf{Redshift}} \\
        & \multicolumn{2}{c|}{\textbf{(AUROC)}} & \multicolumn{2}{c|}{\textbf{(MSE $\times 10^3$)}} & \multicolumn{2}{c|}{\textbf{(AUROC)}} & \textbf{(AUROC)} & \multicolumn{2}{c}{\textbf{(MSE)}} \\
        \cmidrule(lr){2-3} \cmidrule(lr){4-5} \cmidrule(lr){6-7} \cmidrule(lr){8-8} \cmidrule(lr){9-10}
        & \textbf{Lim.} & \textbf{Full} & \textbf{Lim.} & \textbf{Full} & \textbf{Lim.} & \textbf{Full} & \textbf{Full} & \textbf{ZTF} & \textbf{LSST} \\
        \midrule
        Baseline: No Pretraining & 0.637 & 0.853 & 6.02 & 3.85 & 0.498 & 0.527 & 0.901 & 0.079 & 0.289 \\
                       & \scriptsize{±0.005} & \scriptsize{±0.013} & \scriptsize{±0.05} & \scriptsize{±0.31} & \scriptsize{±0.013} & \scriptsize{±0.027} & \scriptsize{±0.003} & \scriptsize{±0.022} & \scriptsize{±0.018} \\
        \midrule
        Classifier (ZTF only) & 0.875 & 0.904 & 4.91 & 3.87 & 0.605 & 0.596 & \multirow{2}{*}{---} & 0.028 & 0.252 \\
                         & \scriptsize{±0.020} & \scriptsize{±0.012} & \scriptsize{±0.10} & \scriptsize{±0.36} & \scriptsize{±0.025} & \scriptsize{±0.008} & & \scriptsize{±0.003} & \scriptsize{±0.004} \\
        \midrule
        Classifier (ZTF + LSST) & 0.879 & 0.910 & \textbf{4.79} & 3.82 & \textbf{0.622} & \textbf{0.616} & \textbf{0.968} & \textbf{0.026} & \textbf{0.177} \\
                          & \scriptsize{±0.011} & \scriptsize{±0.013} & \scriptsize{±0.06} & \scriptsize{±0.39} & \scriptsize{±0.018} & \scriptsize{±0.036} & \scriptsize{±0.006} & \scriptsize{±0.002} & \scriptsize{±0.008} \\
        \midrule
        Classifier + Contrastive & \textbf{0.886} & \textbf{0.914} & 4.87 & \textbf{3.73} & 0.584 & 0.576 & 0.946 & 0.028 & 0.191 \\
                    & \scriptsize{±0.026} & \scriptsize{±0.005} & \scriptsize{±0.03} & \scriptsize{±0.17} & \scriptsize{±0.018} & \scriptsize{±0.028} & \scriptsize{±0.021} & \scriptsize{±0.002} & \scriptsize{±0.013} \\
        \midrule
        Classifier + Adversarial & 0.844 & 0.853 & 5.20 & 4.19 & 0.559 & 0.546 & 0.925 & 0.030 & 0.197 \\
                    & \scriptsize{±0.016} & \scriptsize{±0.012} & \scriptsize{±0.04} & \scriptsize{±0.42} & \scriptsize{±0.024} & \scriptsize{±0.077} & \scriptsize{±0.015} & \scriptsize{±0.003} & \scriptsize{±0.014} \\
        \bottomrule
    \end{tabular}}
    \caption{Performance comparison across downstream tasks for each pretrained model. We report mean $\pm$ standard deviation from five different random fine-tuning samples. Bold indicates best performance per task. The Classifier (ZTF + LSST), Contrastive, and Adversarial models represent our main contributions. For Kepler classification, ``---'' indicates the model was not evaluated on this cross-domain task.}

    \label{tab:downstream}
\end{table*}

\begin{table}
\centering
\resizebox{0.95\linewidth}{!}{
\begin{tabular}{c|l|cc|c}
\toprule
\multirow{2}{*}{} & \multirow{2}{*}{\textbf{Model}} & \multicolumn{2}{c|}{\textbf{Redshifting Data}} & \multirow{2}{*}{\textbf{LSST MSE}} \\
 & & \textbf{ZTF} & \textbf{LSST} & \\ \hline
\multirow{3}{*}{\rotatebox[origin=c]{90}{\textbf{Direct}}} 
& No Pretraining & No & Yes & $0.0750 \pm 0.0252$ \\
& No Pretraining & Yes & Yes & $0.0614 \pm 0.0033$ \\
& Classifier & Yes & Yes & $0.0579 \pm 0.0061$ \\ \hline
\multirow{5}{*}{\rotatebox[origin=c]{90}{\textbf{Zero-shot}}} 
& No Pretraining & Yes & No & $0.1869 \pm 0.0127$ \\
& Classifier & Yes & No & $0.1035 \pm 0.0081$ \\
& Classifier + Contrastive & Yes & No & \textbf{0.0727} $\pm$ \textbf{0.0056} \\
& Classifier + Adversarial & Yes & No & \textbf{0.0744} $\pm$ \textbf{0.0063} \\
& Classifier + Contrastive kNN & Yes & No & $0.0854 \pm 0.0040$ \\
\bottomrule
\end{tabular}}
\caption{Zero-shot redshift estimation performance on LSST simulations. Direct training methods use LSST labels during training, while zero-shot methods are trained only on ZTF data and evaluated on LSST without any LSST training redshifts. Bold indicates best zero-shot performance. Domain-agnostic models (Contrastive, Adversarial) achieve performance comparable to direct training methods.}
\label{tab:zero-shot}
\vspace{-2em}
\end{table}

\subsection{Simulation-Based Pretraining Effectiveness}
\label{sec:real_downstream}

\textbf{Substantial Improvements from Simulation Pretraining:} Table \ref{tab:downstream} shows that the pretrained models consistently outperform no-pretraining baselines across all downstream tasks. Most remarkably, our pretrained models fine-tuned with only 512 labeled examples (Limited scenario) outperform baseline models trained on the full dataset (3747 labeled transients) for both classification and anomaly detection tasks on real data. 

\textbf{Cross-Domain Generalization:} Our models demonstrate impressive generalization capabilities, successfully transferring from explosive transient phenomena to stellar variability. On Kepler variable star classification, the best pretrained classifier achieves $0.968 \pm 0.018$ AUROC compared to $0.901 \pm 0.003$ without pretraining. This cross-domain transfer is particularly remarkable given the distinct observational characteristics of space-based Kepler versus ground-based ZTF/LSST surveys and the fundamental differences between explosive transients and variable stars. The substantial performance improvement on very different data demonstrates an unexpected emergent capability of our pretraining method.

\subsection{Zero-Shot Transfer Performance}
Table \ref{tab:zero-shot} reveals that the additional contrastive and adversarial objectives provide their most significant advantages in zero-shot scenarios. Most remarkably, our contrastive zero-shot approach achieves $0.0727 \pm 0.0056$ MSE, actually outperforming a model trained directly on LSST redshift data ($0.0750 \pm 0.0252$ MSE) and approaching the performance of models trained on both ZTF and LSST redshifts ($0.0614 \pm 0.0033$ MSE).

This performance demonstrates a clear trade-off between fine-tuning and zero-shot capabilities. While the standard classifier outperforms contrastive and adversarial variants on supervised fine-tuning tasks (Table \ref{tab:downstream}), it shows substantially degraded zero-shot performance ($0.1035 \pm 0.0081$ MSE versus $0.0727 \pm 0.0056$ MSE for contrastive). These results indicate that explicit cross-survey alignment through domain-agnostic objectives is essential for effective zero-shot transfer but may introduce constraints that limit performance when target-domain fine-tuning data is available.

\section{Conclusion}

We present a simulation-based pretraining approach that addresses the scarcity of labeled astronomical data by leveraging synthetic light curves from multiple surveys. Our models demonstrate substantial improvements over training from scratch, with pretrained models using only 512 real labeled examples significantly outperforming baselines trained on full datasets with 7$\times$ more data for classification and anomaly detection tasks.

Key findings include: (1) effective cross-domain transfer from explosive transients (ZTF and LSST ground-based surveys) to stellar variability (Kepler space telescope) despite fundamental astrophysical and instrumental differences, revealing unexpected emergent capabilities, and (2) domain-agnostic training objectives that enable zero-shot performance matching models trained directly on target survey data, with our contrastive model achieving comparable LSST redshift performance without any LSST training data.

These findings have immediate relevance for LSST beginning operations in late 2025. Our models can be deployed from day one without extensive domain-specific labeling, dramatically accelerating early science returns. The demonstrated improvement in data efficiency addresses the perennial challenge of limited expert annotations in astronomy. Our approach provides a practical framework for scientific domains where simulations encode substantial domain expertise but labeled observational data remains scarce.

\section*{Acknowledgements}

This work used Bridges-2 at Pittsburgh Supercomputing Center through allocation PHY240105 from the Advanced Cyberinfrastructure Coordination Ecosystem: Services \& Support (ACCESS) program \citep{NSF-ACCESS-Boerner2023}, which is supported by U.S. National Science Foundation grants \#2138259, \#2138286, \#2138307, \#2137603, and \#2138296.

This work made use of the \texttt{python} programming language and the following packages: \texttt{numpy} \citep{numpy}, \texttt{matplotlib} \citep{matplotlib}, \texttt{scikit-learn} \citep{scikit-learn}, \texttt{pandas} \citep{pandas}, \texttt{astropy} \citep{astropy}, and \texttt{pytorch} \citep{paszke2019pytorchimperativestylehighperformance}.



\bibliography{main}
\bibliographystyle{icml2025}

\newpage
\appendix




\section{Model and Training Details} 
\label{sec:model_info}
\subsection{Classifier Training}
We enable cross-survey compatibility by using a survey-agnostic input $[\lambda_p, t_i, f_i]$ format described in Section~\ref{sec:datasets}. This input method \citep{huang2023predicting, gupta2024, gupta2025} specifically allows for the usage of the same model across surveys, something not facilitated by many previous input methods.

Training uses categorical cross-entropy loss with early stopping when validation loss plateaus for 5 epochs. We extract representations from the penultimate layer (128-dimensional) as our learned embeddings $z_i = C_L(X_i)$ for downstream tasks.

\subsection{Adversarial Training Algorithm}

\label{sec:adv_training}

Our adversarial pretraining is summarized in Algorithm \ref{alg:adversarial_training}. Here, \( \bm{X}_i \) denotes the input light curve, \( c_i \) is its class label, and \( O_i \in \{\text{ZTF}, \text{LSST}\} \) indicates the observatory. The latent representation is extracted as \( \bm{z}_i = C_L(\bm{X}_i) \), where \( C_L \) is the penultimate layer of the classifier. The categorical cross-entropy loss is denoted by \( H(p, q) \), where \( p \) is a predicted distribution and \( q \) is a target one-hot vector.

\begin{algorithm}
\caption{Adversarial Training}
\label{alg:adversarial_training}
\begin{algorithmic}[1]
\REQUIRE Dataset \( \{(\bm{X}_i, c_i, O_i)\}_{i=1}^N \): light curves, class labels, and observatory labels
\REQUIRE Classifier \( C \), Discriminator \( D \)
\STATE Initialize \( C \) and \( D \) with random weights
\REPEAT
    \STATE \textbf{// Step 1: Train the discriminator}
    \STATE Freeze the classifier \( C \)
    \STATE For each sample, compute latent representation \( \bm{z}_i = C_L(\bm{X}_i) \)
    \STATE Compute discriminator loss:
    \[
        \mathcal{L}_D = H(D(\bm{z}_i), O_i)
    \]
    \STATE Update \( D \) to minimize \( \mathcal{L}_D \)
    \STATE \textbf{// Step 2: Train the classifier}
    \STATE Freeze the discriminator \( D \)
    \STATE Compute classifier loss with adversarial objective:
    \[
        \mathcal{L}_C = H(C(\bm{X}_i), c_i) - H(D(C_L(\bm{X}_i)), O_i)
    \]
    \STATE Update \( C \) to minimize \( \mathcal{L}_C \)
\UNTIL convergence
\end{algorithmic}
\end{algorithm}

\subsection{Contrastive Training Algorithm}
\label{sec:con_training}

For our supervised contrastive loss, we use the contrastive objective proposed in \citet{SimCLR} for model pretraining. It is formally defined as follows:

\begin{equation}
    \ell_{i,j} = -\log \frac{\exp\left(\text{sim}\left(\bm{z}_i, \bm{z}_j\right)/\tau\right)}{\sum_{k=1}^{2N} \mathbbm{1}_{[k \neq i]} \exp\left(\text{sim}\left(\bm{z}_i, \bm{z}_k\right)/\tau\right)}
\end{equation}

where:
\( \bm{z}_i, \bm{z}_j \) are the latent representations from the classifier, \( \text{sim}(\bm{z}_i, \bm{z}_j) \) denotes cosine similarity: \( \text{sim}(\mathbf{a}, \mathbf{b}) = \frac{\mathbf{a} \cdot \mathbf{b}}{\|\mathbf{a}\| \|\mathbf{b}\|} \), \( \tau > 0 \) is a temperature parameter that scales the similarity scores, and the loss is computed for all pairs \( (i, j) \) where \( \bm{X}_i \) and \( \bm{X}_j \) share the same class label.

The total supervised contrastive loss is computed by summing \( \ell_{i,j} \) over all valid positive pairs in a batch. This encourages latent vectors from the same class to be close together, while implicitly pushing apart representations from other classes.

Our contrastive pretraining is summarized in Algorithm \ref{alg:scl_training}. We set $\tau=0.5$ similar to the default set in previous work \citep{SimCLR}.

\begin{algorithm}[H]
\caption{Supervised Contrastive Training}
\label{alg:scl_training}
\begin{algorithmic}[1]
\REQUIRE Training set \( \{(\bm{X}_i, c_i)\}_{i=1}^N \); light curves, class labels
\REQUIRE Classifier \( C \), temperature parameter \( \tau \)
\STATE Initialize \( C \) with random weights
\REPEAT
    \STATE Compute classification loss
    \[
        \mathcal{L}_C = H(C(\bm{X}_i), c_i)
    \] 
    \STATE Compute latent representations \( \bm{z}_i = C_L(\bm{X}_i) \)
    \STATE Initialize total contrastive loss \( \mathcal{L}_{\text{SCL}} \leftarrow 0 \), counter \( M \leftarrow 0 \)
    \FOR{each anchor sample \( i \in \{1, \dots, N\} \)}
        \STATE Let \( P(i) = \{ j \neq i : c_j = c_i \} \)
        \FOR{each \( j \in P(i) \)}
            \STATE Compute pairwise contrastive loss:
            \[
                \ell_{i,j} = -\log \frac{\exp\left(\text{sim}\left(\bm{z}_i, \bm{z}_j\right)/\tau\right)}{\sum_{k \neq i} \exp\left(\text{sim}\left(\bm{z}_i, \bm{z}_k\right)/\tau\right)}
            \]
            \STATE Accumulate loss: \( \mathcal{L}_{\text{SCL}} \leftarrow \mathcal{L}_{\text{SCL}} + \ell_{i,j} \)
            \STATE Increment counter: \( M \leftarrow M + 1 \)
        \ENDFOR
    \ENDFOR
    \STATE Compute mean contrastive loss: \( \mathcal{L}_{\text{SCL}} \leftarrow \mathcal{L}_{\text{SCL}} / M \)
    \STATE Compute total loss: \( \mathcal{L} \leftarrow \mathcal{L}_{\text{SCL}} + \mathcal{L}_C \)
    \STATE Update \( C \) to minimize \( \mathcal{L} \)
\UNTIL convergence
\end{algorithmic}
\end{algorithm}

    

\subsection{Fine-Tuning}

\label{sec:fine-tuning}

\subsubsection{General Tasks}
For in-domain tasks (ZTF/LSST), we freeze the pretrained encoder and fine-tune only the task-specific MLP head. For cross-domain tasks (Kepler), we unfreeze the initial GRU layers to enable domain adaptation while preserving core learned representations. These is motivated by previous transfer learning research \citep{gupta2025}.

\subsubsection{Anomaly Detection}
Following \citep{gupta2024}, we use a classifier-based approach where a fine-tuning classifier is trained on a set of \textit{normal} data. The penultimate layer of this classifier is then used as a latent space for anomaly detection and an isolation forest \citep{isolationforest} is trained using this latent space to detect anomalies. This method has state-of-the-art (SoTA) performance for anomaly detection on real data.

\subsection{Architecture and Training}
We implement a GRU-based encoder \citep{GRU} with 100 hidden units and two fully-connected layers, producing 128-dimensional latent representations. GRUs provide superior efficiency compared to LSTMs while outperforming standard RNNs \citep{GRUvsLSTM}, making them well-suited for astronomical time series \citep{avacado, gupta2024, Muthukrishna19RAPID}.

Training uses the Adam optimizer \citep{adam} with early stopping after 5 epochs without validation improvement. Convergence times on V100 GPUs: 10 minutes (Classifier), 20 minutes (Adversarial), 45 minutes (Contrastive). The final experiments required an estimated 15-25 GPU hours, however ideation and experimentation required considerably more.





\begin{figure*}[ht]
    \centering
    \begin{tabular}{ccc}
        \includegraphics[width=0.3\linewidth]{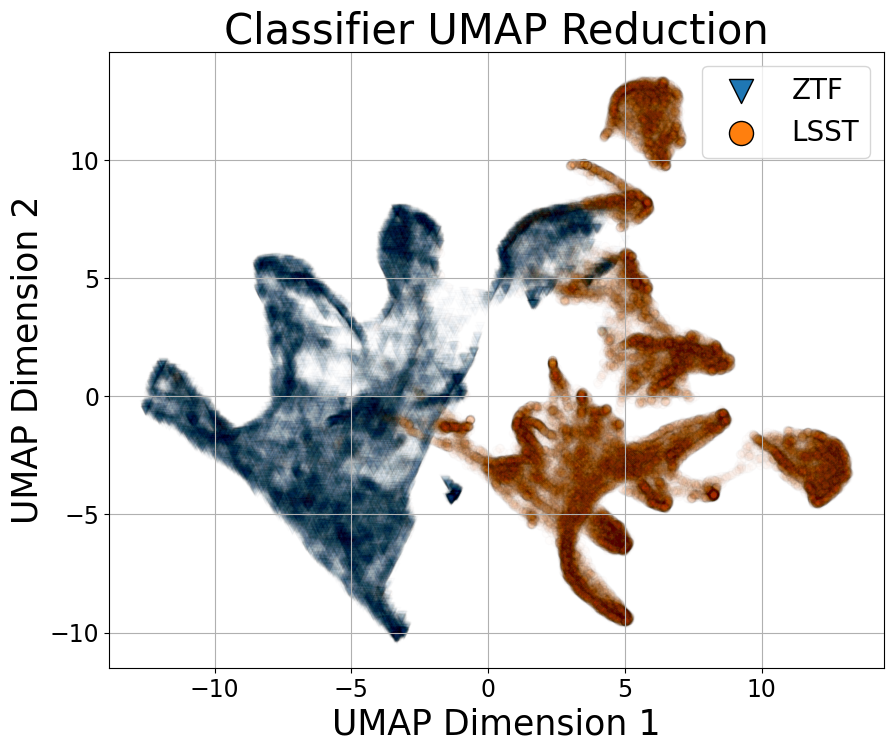} 
        &
        \includegraphics[width=0.3\linewidth]{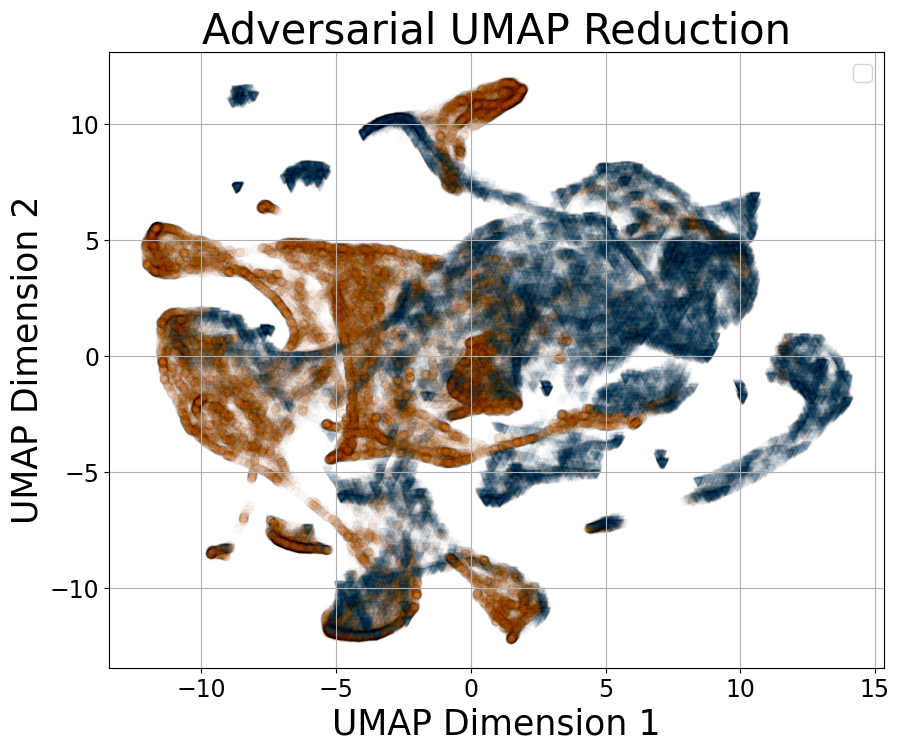}
        &
        \includegraphics[width=0.3\linewidth]{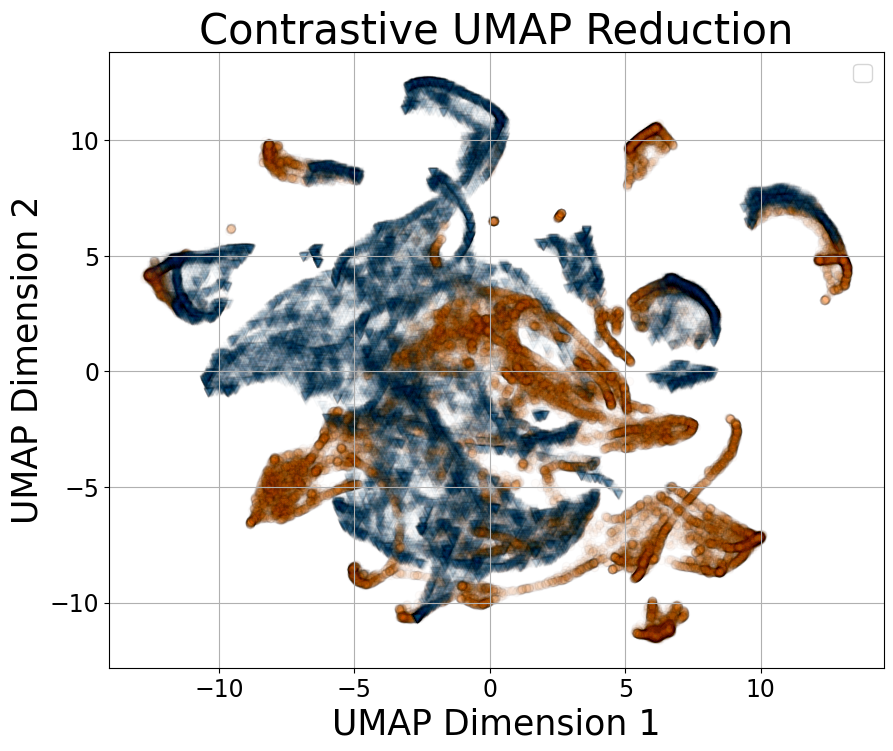} \\
        Classifier & Adversarial & Contrastive
    \end{tabular}
    \caption{UMAP visualization of learned representations across all classes. The standard classifier maintains distinct ZTF (brown) and LSST (blue) clusters, while domain-agnostic training methods appear to unify surveys within each astronomical class slightly better.}
    \label{fig:umap}
\end{figure*}

\begin{figure*}[ht]
    \centering
    \begin{tabular}{ccc}
        \includegraphics[width=0.3\linewidth]{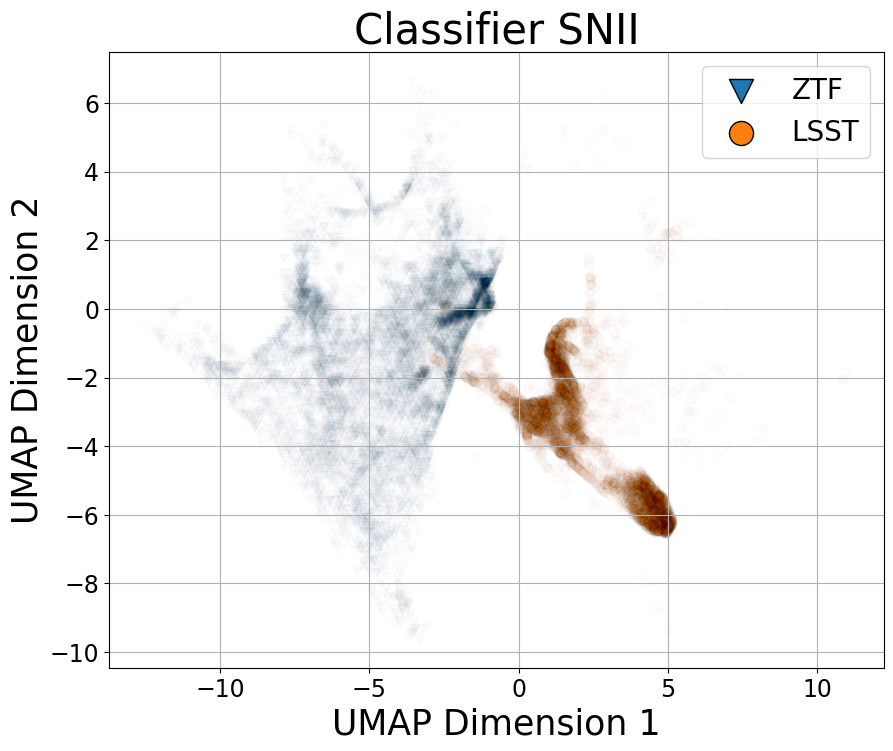} 
        &
        \includegraphics[width=0.3\linewidth]{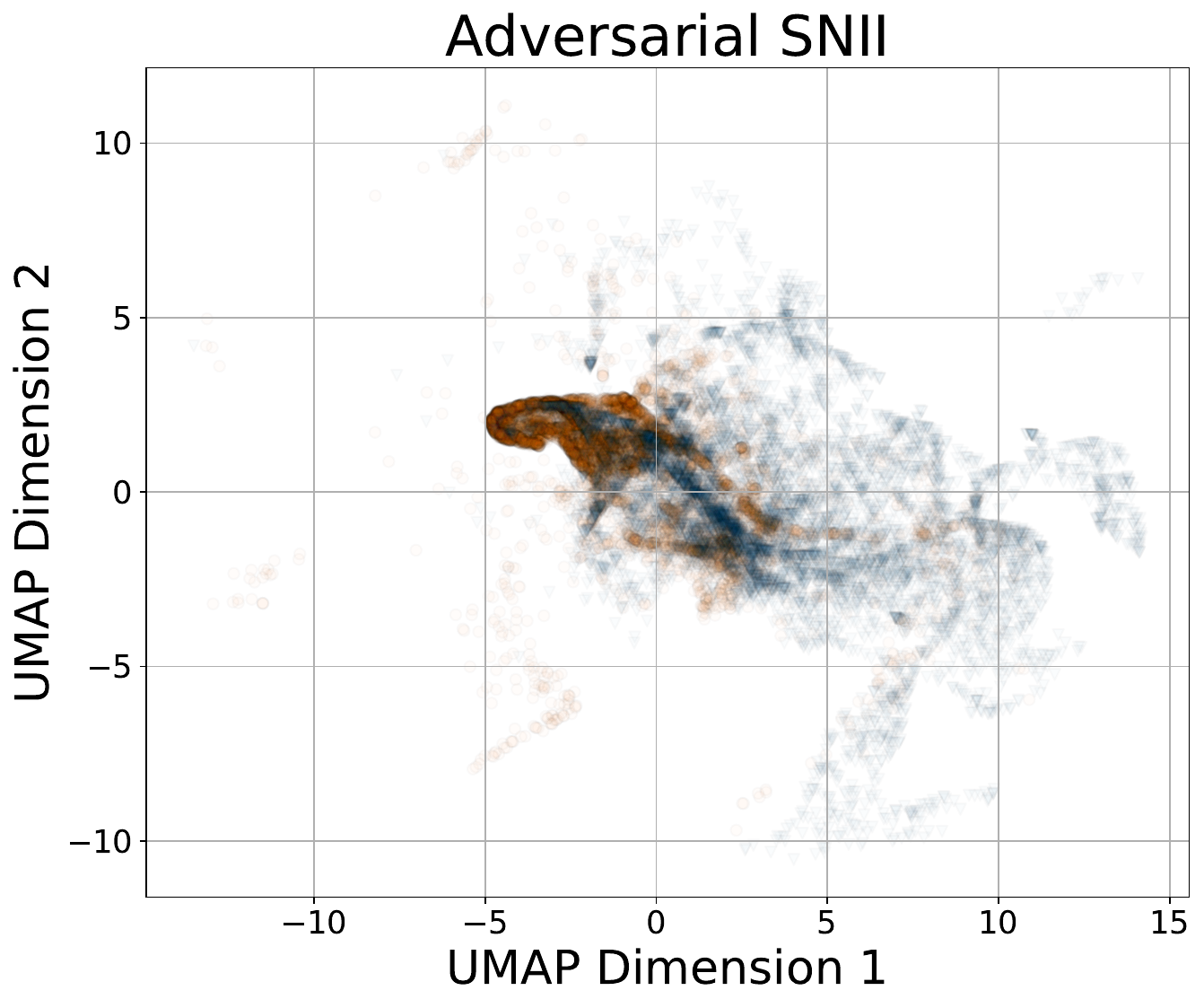}
        &
        \includegraphics[width=0.3\linewidth]{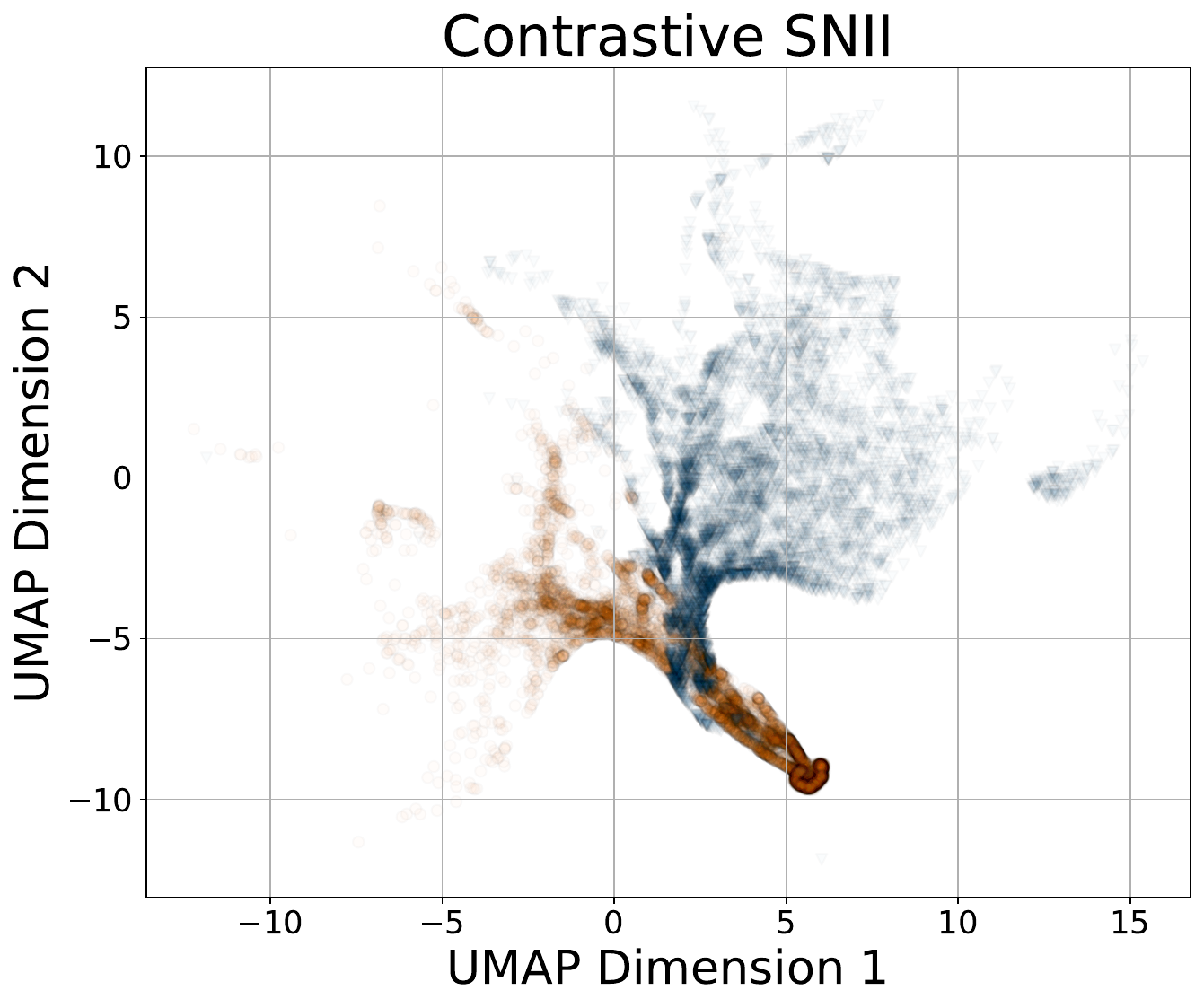} \\
        Classifier & Adversarial & Contrastive
    \end{tabular}
    \caption{Class-level alignment for Type II supernovae only. Domain-agnostic methods achieve better survey overlap in the latent representations while preserving astronomical class structure, indicating at possible cross-survey transfer capabilities.}
    \label{fig:umapclass}
\end{figure*}

\section{Qualitative Latent Space Analysis}
\label{sec:umap-latent}

We visualize the learned representations using UMAP \citep{UMAP} to assess domain alignment quality. Figure~\ref{fig:umap} shows that standard classifiers maintain distinct ZTF and LSST clusters, indicating survey-specific artifacts in the learned embeddings. In contrast, both adversarial and contrastive training appear to merge these embedding, creating slightly more unified representations where survey identity becomes secondary to astronomical class. Future work should explore this claim quantitatively.

Figure~\ref{fig:umapclass} demonstrates this alignment at the class level, showing Type II supernovae from both surveys occupying the same latent region. This class-level unification may validate that our domain-agnostic objectives preserve astronomical semantics while eliminating instrumental biases - precisely the behavior required for effective zero-shot transfer.

Fig.~\ref{fig:umap} and \ref{fig:umapclass} show a UMAP \citep{UMAP} visualization of the penultimate layer of our neural network classifiers. As seen, contrastive and adversarial models help unify the distributions of ZTF and LSST in the latent space down to the class level.

\section{Dataset Details}

\subsection{ZTF and LSST simulations}
Table~\ref{table:simulated} provides the detailed class distribution for our simulated pretraining datasets. The redshift distribution is illustrated in Figure~\ref{fig:redshift_distributions}.

\subsubsection{Zero-Shot Redshift Range Restriction}
We implement a redshift cut ($\text{z} < 0.4$) for zero-shot LSST evaluation to match the ZTF training distribution and to prevent distribution shift. This reduces the LSST evaluation set from 9,990 to 2,596 objects but ensures that performance differences reflect instrumental transfer rather than extrapolation failure.

LSST will observe transients up to z $\sim$ 1.2 (Figure~\ref{fig:redshift_distributions}), but zero-shot models trained only on ZTF data ($\text{z} < 0.4$) cannot reliably predict redshifts beyond their training range. Evaluating on the full LSST redshift range would conflate two distinct problems: instrumental differences between surveys and extrapolation to unseen redshift regimes.  For higher-redshift LSST objects, we perform survey-specific fine-tuning (Table~\ref{tab:downstream}) for each pretraining approach.

\subsection{ZTF Real Data}
\begin{figure*}
    \centering
    \begin{tabular}{cc}
        \includegraphics[width=0.64\linewidth]{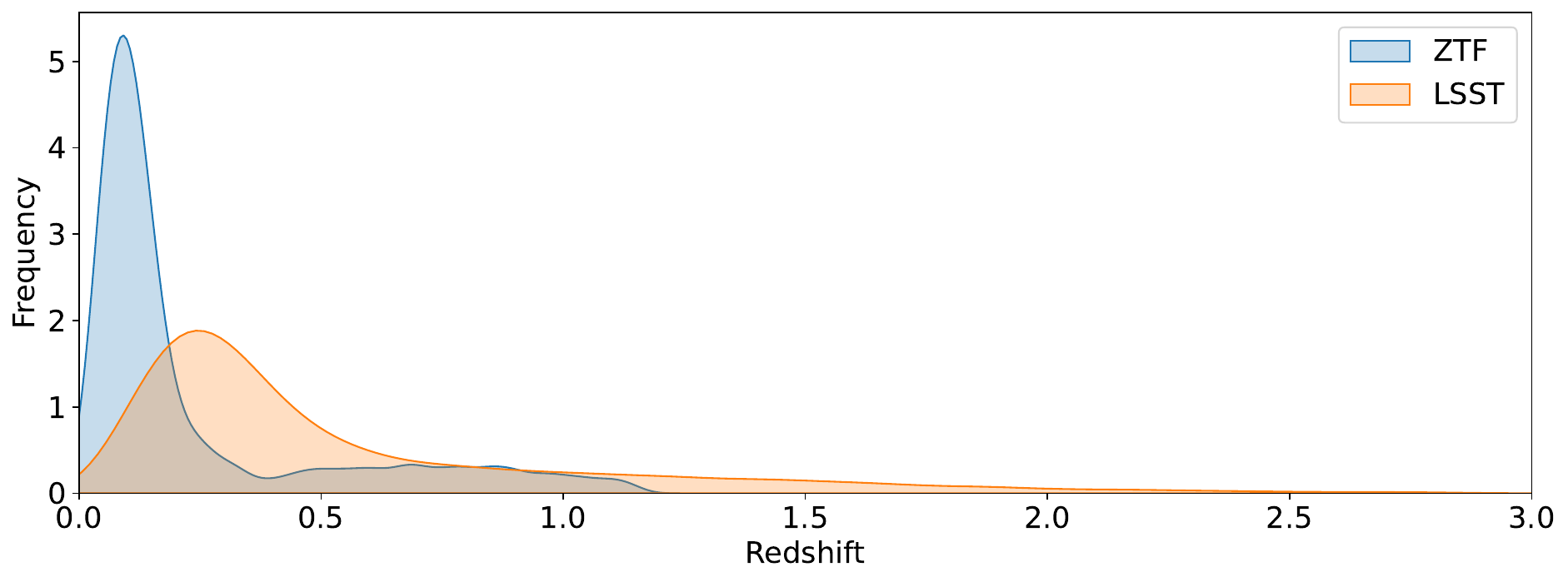} 
        &
        \includegraphics[width=0.36\linewidth]{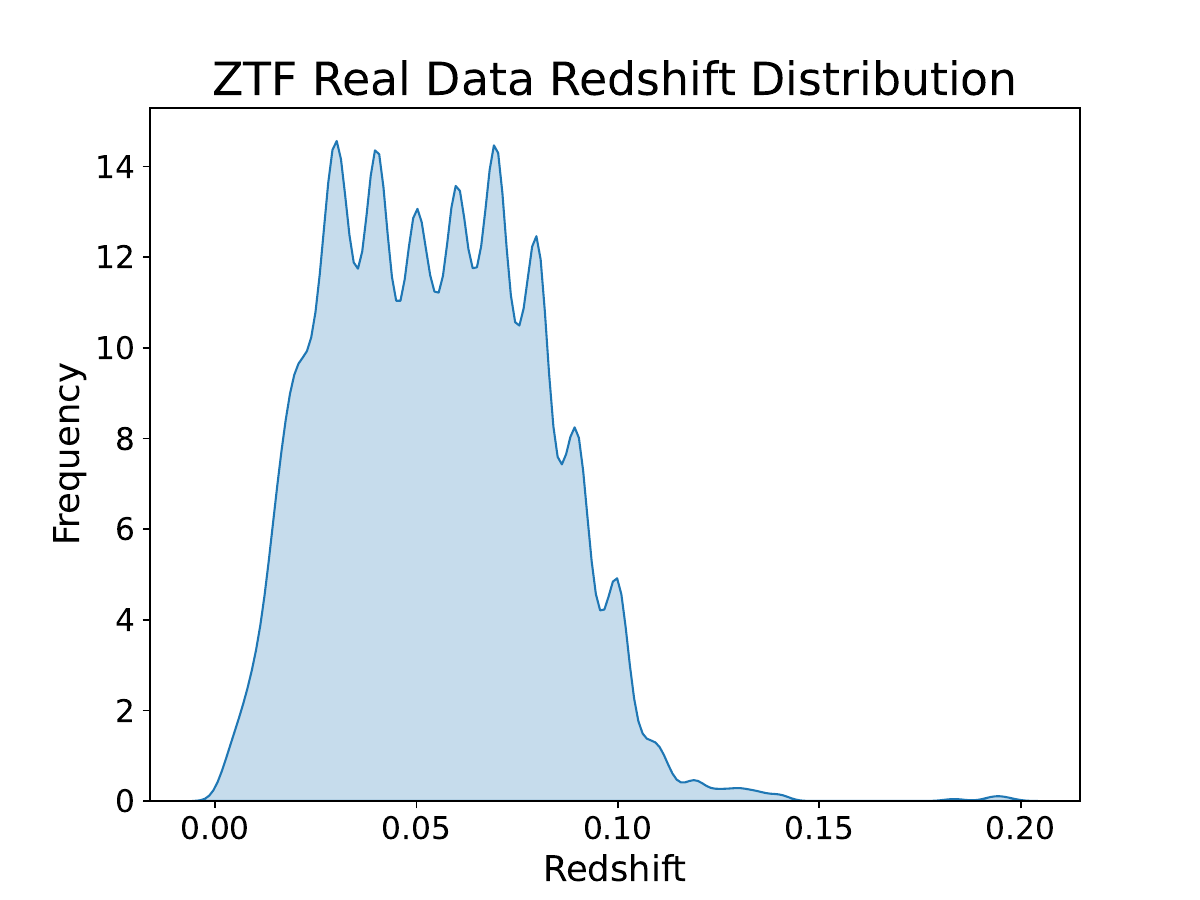} \\
        (a) Simulated ZTF and LSST & (b) Real ZTF observations
    \end{tabular}
    \caption{Redshift distributions across datasets. (a) Simulated ZTF transients span $z = 0.0-0.4$ while LSST simulations extend to $z \sim 3.0$, reflecting LSST's deeper photometric sensitivity ($r \sim 25$ mag) compared to ZTF ($r \sim 21$ mag). (b) Real ZTF observations show a similar distribution to ZTF simulations, validating the simulation fidelity. For zero-shot evaluation, we restrict LSST objects to ZTF's redshift range ($z < 0.4$) to test transfer learning within the training distribution.}
    \label{fig:redshift_distributions}
\end{figure*}

Table~\ref{table:bts_distribution} details the class distribution for our real ZTF evaluation dataset from the Bright Transient Survey (BTS). All objects have spectroscopic confirmation with the redshift distribution illustrated in Figure~\ref{fig:redshift_distributions}. The redshift distribution is much lower as the BTS preferences classifying brighter, and hence lower redshift transients. We do not use any real light curves for pretraining and reserve them solely for evaluation

For anomaly detection experiments, we designate the following rare transient classes as anomalies: Tidal Disruption Events (TDE), Calcium-rich transients, Intermediate Luminosity Red Transients (ILRT), Luminous Blue Variables (LBV), Luminous Red Novae (LRN), Super-Luminous Supernovae Types I and II (SLSN-I, SLSN-II), peculiar Type Ia subtypes (SN Ia-91T, SN Ia-91bg), Type Ibn supernovae (SN Ibn), Broad-Line Type Ic supernovae (SN Ic-BL), and Type Icn supernovae (SN Icn). These classes represent $< 1$\% of our real transient dataset and were selected as anomalies because of their low observation rates.

\subsection{Kepler Variable Stars}
We evaluate cross-domain generalization using 9,501 variable star light curves from the Kepler mission \citep{Koch2010,Borucki2010}, processed by \citet{Audenaert2021}. The dataset spans eight stellar variability classes from aperiodic variables to regular pulsators (Table~\ref{table:kepler}).

To ensure compatibility with our transient survey pretraining data, we downsample the high-cadence Kepler observations from 1024 measurements (30-minute sampling) to $\sim$205 measurements by averaging every five consecutive observations. This preprocessing matches the typical observation count in ZTF and LSST light curves while preserving the essential variability signatures needed for classification.

This cross-domain evaluation tests whether models pretrained on explosive transients can generalize to fundamentally different astrophysical phenomena—stellar pulsations and eclipsing systems—despite having no exposure to periodic variability during training.

\begin{table*}
\begin{center}
\resizebox{0.9\textwidth}{!}{
\begin{tabular}{@{}l@{\hspace{12pt}}|@{\hspace{8pt}}c@{\hspace{8pt}}c@{\hspace{8pt}}c@{\hspace{8pt}}c@{\hspace{8pt}}c@{\hspace{8pt}}c@{\hspace{8pt}}c@{\hspace{8pt}}c@{\hspace{8pt}}|@{\hspace{8pt}}c@{}}
\toprule
\multirow{2}{*}{\textbf{Dataset}} & \multicolumn{7}{c}{\textbf{Astronomical Transient Classes}} & & \multirow{2}{*}{\textbf{Total}} \\
\cmidrule{2-8}
& \textbf{SNIa} & \textbf{SNIa-91bg} & \textbf{SNIax} & \textbf{SNIb/c} & \textbf{SNII} & \textbf{TDE} & \textbf{SLSN-I} & \textbf{AGN} & \\
\midrule
\multicolumn{10}{@{}l@{}}{\textit{Pretraining Data (Classification)}} \\
ZTF Simulations   & 9,436 & 10,663 & 10,681 & 6,769  & 31,193 & 9,260 & 10,451 & 8,627 & 87,080 \\
LSST Simulations   & 8,427 & 6,079  & 8,298  & 7,664 & 9,465 & 9,686 & 6,947 & 7,822 & 64,388 \\ 
\midrule
\multicolumn{10}{@{}l@{}}{\textit{Downstream Task Data (Redshift Estimation)}} \\
ZTF Simulations & 967 & 1,140 & 1,124 & 702  & 3,213 & 932 & 1,053 & 869 & 10,000\\
LSST Simulations    & 1,236  & 1,238  & 1,226  & 1,286  & 1,248  & 1,279  & 1,253  & 1,234 & 9,990  \\
\quad Zero-Shot LSST Subset & 265 & 273 & 535 & 258 & 333 & 215 & 446 & 271 & 2,596 \\
\bottomrule
\end{tabular}}
\caption{Class distribution of simulated astronomical transients. Classification data is used for pretraining, while redshift data (completely separate objects) tests downstream transfer learning. The zero-shot subset contains LSST objects within ZTF's redshift range ($z < 0.4$) for realistic evaluation.}
\label{table:simulated}
\end{center}
\end{table*}

\begin{table*}
\begin{center}
\begin{tabular}{@{}l@{\hspace{8pt}}|@{\hspace{6pt}}c@{\hspace{6pt}}c@{\hspace{6pt}}c@{\hspace{6pt}}c@{\hspace{6pt}}|@{\hspace{6pt}}c@{}}
\toprule
\textbf{Task} & \textbf{SNIa} & \textbf{SNIb/c} & \textbf{SNII} & \textbf{Anomaly} & \textbf{Total} \\
\midrule
Classification & 771 (107) & 2,828 (350) & 148 (12) & 0 & 3,747  \\
Redshift Estimation & 771 (107) & 2,828 (350) & 148 (12) & 0 & 3,747 \\
Anomaly Detection & 771 (107) & 2,828 (350) & 148 (12) & 0 (38) & 3,785 \\
\bottomrule
\end{tabular}
\caption{Real ZTF transient data distribution across downstream tasks. Numbers in parentheses indicate evaluation set sizes. Limited scenarios use 512 randomly selected objects per task from the training portion. All tasks share the same underlying object pool from the ZTF Bright Transient Survey.}
\label{table:bts_distribution}
\end{center}
\end{table*}

\begin{table*}
\begin{center}
\resizebox{0.95\linewidth}{!}{
\begin{tabular}{@{}l@{\hspace{8pt}}|@{\hspace{6pt}}c@{\hspace{6pt}}c@{\hspace{6pt}}c@{\hspace{6pt}}c@{\hspace{6pt}}c@{\hspace{6pt}}c@{\hspace{6pt}}c@{\hspace{6pt}}c@{\hspace{6pt}}c@{\hspace{6pt}}|@{\hspace{6pt}}c@{}}
\toprule
\textbf{Task} & \textbf{Aperiodic} & \textbf{Constant} & \textbf{Contact} & \textbf{DSCT/BCEP} & \textbf{Eclipse} & \textbf{GDOR/SPB} & \textbf{Instr} & \textbf{RR/CEP} & \textbf{Solar} & \textbf{Total} \\
\midrule
Kepler Classification & 831 & 1,000 & 2,260 & 772 & 974 & 630 & 1,171 & 63 & 1,800 & 9,501 \\
\bottomrule
\end{tabular}}
\caption{Distribution of Kepler variable star data across nine stellar variability classes used for cross-domain evaluation (aperiodic, constant, contact binaries, $\delta$ Scuti / $\beta$ Cephei, eclipsing binaries, $\gamma$ Doradus(GDOR) / Slowly Pulsating B stars (SPB), instrument variability, RR Lyrae / Cepheids, Solar-like oscillators). This dataset tests generalization from explosive transients (pretraining) to variable star phenomena, representing a fundamentally different astronomical domain with space-based, high-cadence observations.}
\label{table:kepler}
\end{center}
\end{table*}

\section{Further Analysis}

\label{sec:further-analysis}

\subsection{Adversarial vs. Contrastive Loss}

Incorporating the specialized training techniques proposed in this work does not improve model performance on supervised downstream tasks (Table~\ref{tab:downstream}), which is expected since these loss functions are designed for cross-survey alignment rather than task-specific optimization. Adversarial training shows a slight performance decrease compared to standard classifiers. We attribute this to the adversarial netowrk effectively placing a penalty term for domain alignment, whereas the contrastive model receives explicit supervision through class-based positive pairs.

On zero-shot tasks, both adversarial and contrastive models substantially outperform standard classifiers (Table~\ref{tab:zero-shot}), demonstrating that the domain-agnostic embedding enables meaningful cross-survey transfer. We note that training with both contrastive and adversarial losses simultaneously does not improve performance in comparison to a purely contrastive model.

\subsection{Performance on Simulations}

\label{sec:gap-real-sims}

Fine-tuned models consistently underperform on real data compared to simulations, highlighting the domain gap between synthetic and observational data. For redshift estimation, our best model achieves  $R^2$ score of $0.431 \pm 0.053$ on real ZTF data versus $0.580 \pm 0.011$ on simulations. Models can leverage these physics-based simulations as effective starting points but still require labeled real data to perform well. This gap between real data and simulations is why there is a significant performance gap between the \texttt{Limited} and \texttt{Full} evaluation scenarios. In other words, models trained on simulations need to be fine-tuned on real data to work well.

\subsection{Anomaly Detection}
Anomaly detection is the only task in which we do not see a performance improvement from the \texttt{Limited} to \texttt{Full} settings (Table \ref{tab:downstream}). By definition, anomalies are objects that astronomers find interesting. By using human-defined simulations to pretrain, they are naturally equipped to detect specifically what humans find interesting. The gap between simulations and real data is what anomaly detection pipelines are trying to fill. Further analysis of anomaly detection specifically is out of the scope of this work, however we hope that future researchers analyze the nature of this task and how domain expertise can be incorporated into it.

\section{kNN Zero-Shot Estimation}
\label{sec:knn}

Aside from fine-tuning an MLP, we also evaluate using a k Nearest Neighbors approach for zero-shot estimation \citep{maven, AstroCLIP}. To perform zero-shot redshift estimation for an LSST object, we first find the $k = 100$ closest ZTF embeddings to the LSST light curve embedding in the latent space. Then, we use the distance-weighted average of the corresponding redshifts to estimate the final redshift of the LSST object. As seen in Table \ref{tab:zero-shot}, this zero-shot estimation method performs worse than using a directly trained MLP.


\end{document}